\documentclass[aps,pra,superscriptaddress,amsmath,amssymb,preprintnumbers
,twocolumn,showpacs]{revtex4}

\usepackage{graphicx}

\usepackage{dcolumn}

\usepackage{bm}

\usepackage{amsmath}

\begin{document}


\title{\LARGE \bf Interpreting concurrence in terms of covariances in a
generalized spin star system }

\author{F. Palumbo\footnote[2]{The authors AN and AM dedicate
this paper to the memory of Francesca Palumbo, recently passed}}

\author{A. Napoli}

\author{A. Messina}

\affiliation{ MIUR and Dipartimento di Scienze Fisiche ed
Astronomiche,
via Archirafi 36, 90123 Palermo, Italy, \\
E-mail:
messina@fisica.unipa.it }

\begin{abstract}
The quantum dynamics of $M$ pairwise coupled spin $\frac{1}{2}$ is
analyzed and the time evolution of the entanglement get
established within a prefixed couple of spins is studied. A
conceptual and quantitative link between the concurrence function
and measurable quantities is brought to light providing a physical
interpretation for the concurrence itself as well as a way to
measure it. A generalized spin star system is exactly investigated
showing that the entanglement accompanying its rich dynamics is
traceable back to the covariance of appropriate commuting
observables of the two spins.

\end{abstract}


\maketitle

\medskip
\pagebreak
\section{Introduction}
Quite recently a growing attention has been devoted to interacting
spin systems \cite{Loss, Karbach, Dobrovitski, Geza, Cucchietti}
also in view of the fact that they can be successfully used for
gate operations in solid state quantum computation processors
\cite{Karbach, Burkard, Awschalom, Imamoglu}. They indeed provide
scalable systems that can be easily integrated into standard
silicon technology. Spin sets with an assigned one- or
multi-dimensional spatial distribution appear to be comparatively
more promising candidates for the realization of entangled states
in matter systems.

Numerous papers published over the last few years witness the
central role played by spin systems, both from a theoretical and
applicative point of view, in the emerging field of quantum
entanglement in solid state physics. As an example, and in
connection to the subject we are going to discuss in the present
paper, it is worth explicitly citing those papers dealing with
investigation on the entanglement get established within a pair of
spins belonging to a spin ensemble whose dynamics is dominated by
XXX \cite{XXX}, XXZ \cite{XXZ}, XYZ \cite{XYZ1, XYZ2} and XY
\cite{XY1, XY2} Heisenberg interaction models. In these cases the
analysis is developed through the evaluation of the concurrence
function appropriate to estimate the degree of quantum
correlations get established in a two-spin system.

It is important to underline that generally speaking quantifying
and controlling entanglement is a crucial challenge both from a
theoretical point of view and in consideration of its applicative
potentialities in various field of quantum information.
 It is in addition important to stress that studying spin
systems is however of remarkable interest in its own. It is well
known, for example, that a magnetic sample can be suitably
analyzed adopting a general Hamiltonian model describing a system
of $N$ spins coupled by exchange interactions with arbitrary range
and strength \cite{magnetism}.

In this paper we study the quantum dynamics of a system of $M$
Heisenberg pairwise coupled spins $\frac{1}{2}$  investigating in
particular the time evolution of the entanglement get established
within a spin pair subsystem. The analysis here reported enable us
to bring to light a new conceptual and quantitative link between
the concurrence function and an observable quantity having a clear
physical meaning suggesting, at the same time, a simple way of
measuring entanglement in a two spin system without necessarily
reconstructing its state. Our approach focusses on a generalized
spin star system whose dynamics provides an enlightening key to go
deep into the physical meaning of entanglement and concurrence. In
this framework we succeed in giving a simple recipe to control the
ability of the system in developing only classical  or also
quantum correlations.

\section{Heisenberg Interacting spin systems}

Our physical system consists of $M$ two-level objects whose
quantum dynamics is completely described by spin operators
$\vec{S}^{(i)}\equiv 2\vec{\sigma}^{(i)}$ $(i=1,...,M)$,
$\vec{\sigma}^{(i)}$ being the Pauli matrix operator pertaining to
the $i-$th two-level subsystem.  The cartesian components of
$\vec{S}^{(i)}$, $S_x^{(i)},S_y^{(i)}$, and $S_z^{(i)}$, fulfill
the usual angular momentum commutation relations and $\vert \pm
1_i\rangle$ denote the two eigenstates of $\sigma_z^{(i)}$ that is
$\sigma_z^{(i)} \vert \pm 1_i \rangle=\pm \vert \pm 1_i \rangle$.

In this section we wish to keep our presentation general enough to
encompass several possible physical scenarios of interest. Thus we
adopt the assumption that a spin pair $(i,j)$ belonging to our
ensemble of $M$ spins $(i,j=1,2,...,M; \;i\neq j)$ experiences an
Heisenberg exchange-like interaction describable as proportional
to $\vec{S}^{(i)}\cdot \vec{S}^{(i)}$. The hamiltonian model
representing such a physical situation in the interaction picture
with respect to the free hamiltonian
\begin{equation} \label{H0}
H_0=\omega\sum_{i=1}^{M}S_z^{(i)}
\end{equation}
may be cast in the following general form
\begin{equation}\label{generalH}
H_I=\sum_{j>i=1}^{M}{\alpha_{ij}\vec{S}^{(i)} \cdot\vec{S}^{(j)}}
\end{equation}
$\alpha_{ij}$ being a real coupling constant left undetermined at
the moment. Since $H_I$ is a scalar operator with respect to
$\vec{S}=\sum_{i}\vec{S}^{(i)}$ and in addition commutes with
$H_0$, then the $z-$component of $\vec{S}$ as well as $S^2$ are
constants of motion. As a consequence, preparing the $M$ spins in
an eigenstate of $S_z$, the total system evolves in the
correspondent invariant Hilbert subspace. It is not difficult to
convince oneself that the reduced density matrix relative to a
prefixed couple of spins, say the spins $i$ and $j$, when
expressed into the standard two-spin basis
\begin{equation}\label{standard basis}
\{|1_i,1_j\rangle,|1_i,-1_j\rangle,|-1_i,1_j\rangle,|-1_i,-1_j\rangle\}
\end{equation}
assumes a block diagonal form, each block  being biunivocally
singled out by one of the three possible
 eigenvalues of $S_z^{(ij)}$$\equiv(S_z^{(i)}+S_z^{(j)})$.
Stated another way, the reduced density matrix $\rho_{ij}(t)$,
obtained tracing the density operator of the closed system over
the degrees of freedom relative to all the $M$ spins except $i$
and $j$, assumes the quite simple form
\begin{equation}\label{rogenerica}
\rho_{ij}(t) = \left( \begin{array}{cccc}
a(t) & 0 & 0 & 0 \\
0& b(t) & c(t) & 0 \\
0& c^*(t) & d(t)& 0\\
0& 0&  0& e(t)
\end{array} \right)
\end{equation}
when represented in the ordered basis given by eq. (\ref{standard
basis}) \cite{Pratt}. We wish to stress that the possibility of
writing $\rho_{ij}(t)$ as in eq. (\ref{rogenerica}) directly stems
from the initial preparation being instead independent on the
values of the coupling constants set $\{ \alpha_{ij}\}$. On the
contrary, the matrix elements of $\rho_{ij}$ depend on this set
too.

Scope of this section is to prove the existence of dynamical
properties of our system of $M$ spins relying solely on the
structure of the density matrix $\rho_{ij}(t)$ and not on the
specific form  of the functions $a(t)$, $b(t)$,..., $e(t)$
appearing in eq. (\ref{rogenerica}).

This circumstance appears still more remarkable at the light of
the fact that, generally speaking, it is not possible to exactly
solve the dynamics of $M$ interacting spins described by the
hamiltonian model (\ref{generalH}). Thus results obtained just
exploiting the form of the two-spin reduced density operator
$\rho_{ij}(t)$ (\ref{rogenerica}), besides being valid whatever
the coupling constants are, also provide peculiar tools to test
approximate dynamical solutions when we are unable to exactly
solve the system dynamics.

With these considerations in mind, we now focus on a system of two
spins $\frac{1}{2}$ described at a generic time instant $t$ by a
density matrix like (\ref{rogenerica}) without specifying the
analytic expression of the four population functions $a(t)$,
$b(t)$,$d(t)$, $e(t)$ and the coherence function $c(t)$.

Let's begin by observing that in order to guarantee that the
operator given by equation (\ref{rogenerica}) represents indeed a
density matrix, the inequality
\begin{equation}\label{Laundau condition}
\vert c(t)\vert\leq\sqrt{b(t)d(t)}
\end{equation}
must be satisfied at any time instant $t$ \cite{Landau}. It is
easy to demonstrate that this relation directly stems from the
requirement that all the eigenvalues of $\rho_{ij}$ are not
negative.

An interesting property assumed by each density matrix belonging
to the class defined by eq. (\ref{rogenerica}) concerns the
possibility of getting a bridge between the Peres-Horodecki (P-H)
separability condition
 \cite{P-H1,P-H2} and the measure of
entanglement proposed by Wootters \cite{Wootters98}. The P-H
separability criterium claims that the density matrix $\rho$ of a
bipartite system composed by two-level subsystems, is separable if
and only if all the eigenvalues of the matrix $\sigma$ obtained
from $\rho$ transposing with respect to the indices of only one
subsystem, are not negative. In our case such a matrix, built up
from $\rho_{ij}$, may be cast as follows
\begin{equation}\label{sigma density matrix}
\sigma_{ij}(t) = \left( \begin{array}{cccc}
a(t) & 0 & 0 & c^*(t) \\
0& b(t) & 0 & 0 \\
0& 0 & d(t)& 0\\
c(t)& 0&  0& e(t)
\end{array} \right)
\end{equation}
Thus applying the P-H separability criterium, it is easy to
demonstrate that our two-spin density matrix (\ref{rogenerica}) is
separable at a fixed time instant $t$ if and only if the condition
\begin{equation}\label{separability condition}
\vert c(t) \vert \leq \sqrt{a(t)e(t)}
\end{equation}
is fulfilled. On the other hand, when the density operator of two
two-level systems has the simple form of eq. (\ref{rogenerica}) it
is quite straightforward to evaluate the concurrence function
$C_{onc}(t)$, introduced by Wootters as a measure of entanglement
in bipartite system composed by two qubits, getting
\begin{equation}\label{conc}
C_{onc}(t)=Max\{0, 2(\vert c(t) \vert - \sqrt{a(t)e(t)}) \}
\end{equation}

Considering eq. (\ref{conc}) we immediately deduce that at a
generic time instant $t$ the system is characterized by absence of
entanglement if and only if condition (\ref{separability
condition}) is satisfied in accordance with the P-H separability
criterium.

Let's in addition remark that the presence of entanglement at a
time instant $t$ necessarily implies the existence of at least a
couple of operators $A^{(i)}$ and $B^{(j)}$ acting on the
bidimensional Hilbert spaces of the spin $i$ and $j$ respectively,
such that the correlation function
\begin{eqnarray}\label{covAB}
&&\nonumber C_{AB}(t)\equiv Tr\{ \rho_{ij}(t)A^{(i)} B^{(j)} \}\\
&&-Tr_i\{ \rho_{i}(t)A^{(i)}\}Tr_j\{ \rho_{j}(t)B^{(j)}\}
\end{eqnarray}
is different from zero. In eq. (\ref{covAB}) $\rho_{k}$ $(k=i,j)$
is the reduced density matrix of the spin $k$ and $Tr_k$ denotes
the trace with respect to its degrees of freedom. At the light of
the results expressed by eq. (\ref{separability condition}) and
(\ref{conc}) a suitable pair of operators  satisfying eq.
(\ref{covAB}) must at least fulfill the condition of not being
diagonal in the standard basis (\ref{standard basis}). Thus we are
lead to consider the two operators $\sigma_x^{(i)}$ and
$\sigma_x^{(j)}$. Exploiting the form  of $\rho_{ij}(t)$ as given
by eq. (\ref{rogenerica}) it is immediate to demonstrate that
\begin{equation}\label{Cx}
C_{\sigma_x \sigma_x}(t)=2 \Re{[c(t)]}
\end{equation}
since
\begin{equation}\label{s_x medio}
\langle\sigma_x^{(k)}\rangle\equiv Tr_k\{
\rho_{k}(t)\sigma_x^{(k)}\}=0 \;\;\;\; (k=i,j)
\end{equation}
and
\begin{equation}\label{s_x s_x medio}
\langle\sigma_x^{(i)}\sigma_x^{(j)}\rangle\equiv Tr\{
\rho_{ij}(t)\sigma_x^{(i)} \sigma_x^{(j)}\}=2 \Re{ [c(t) ]}
\end{equation}

If instead we analogously consider the two operators
$\sigma_y^{(i)}$ and $\sigma_y^{(j)}$ we obtain
\begin{equation}\label{Cy}
C_{\sigma_y \sigma_y}(t)=2 \Im{[c(t)]}
\end{equation}
A comparison among eqs. (\ref{Cx}), (\ref{Cy}) and (\ref{conc})
brings to light the existence of a new direct link between the
concurrence function and measurable quantities of clear physical
meaning. We may indeed claim that when different from zero the
concurrence function may be expressed as
\begin{equation}\label{conc Cx Cy a e}
C_{onc}(t)= \sqrt{C_{\sigma_x \sigma_x}^2+C_{\sigma_y
\sigma_y}^2}-2\sqrt{a(t)e(t)}
\end{equation}
$a(t)$ and $e(t)$ being the probability of finding both the spins
$i$ and $j$ in their up states or down states respectively. It is
in addition rather remarkable the fact that if $a(t)=0$ and/or
$e(t)=0$ the concurrence as expressed by eq. (\ref{conc}) reduces
to
\begin{equation}\label{conc Cx Cy}
C_{onc}(t)= 2 \sqrt{\Re{[c(t)]}^2+\Im{[c(t)]}^2}\equiv
\sqrt{C_{\sigma_x \sigma_x}^2+C_{\sigma_y \sigma_y}^2}
\end{equation}
so that it is zero only if both the two quantum covariances
$C_{\sigma_x \sigma_x}(t)$ and $C_{\sigma_y \sigma_y}(t)$ vanish.

This elegant formula suggests in a very transparent way that the
covariances between the $x$ and $y$ components of the two spins
are suitable quantities in order to highlight the presence of
entanglement in our two-spin system. It is however important to
emphasize that $C_{\sigma_x \sigma_x}\neq 0$ $(C_{\sigma_y
\sigma_y}\neq 0)$ doesn't necessarily imply in its own that the
system has developed quantum correlations. This aspect will appear
more clear in the next section where we will analyze the exact
dynamics of a specific system whose hamiltonian model is a
particular case of that expressed by equation (\ref{generalH}).
Concluding this section we wish to stress that the quite simple
form of $\rho_{ij}(t)$, given by eq. (\ref{rogenerica}), which
represents the starting point of our analysis, naturally arises in
many physical contexts not necessarily involving spin
$\frac{1}{2}$ systems. In particular the conditions under which
eq. (\ref{conc Cx Cy}) has been derived are verified for example
when the dynamics of two isolated atoms, each in its own
Jaynes-Cummings cavity is studied \cite{cavity1, cavity2,
cavity3}. In this sense, we may claim that the concepts and tools
reported in this paper are flexible enough to be exported into
physical situations more general than the spin system here
envisaged.

\section{Generalized spin star system}
Appropriately choosing the coupling constants $\alpha_{ij}$ the
quite general hamiltonian model (\ref{generalH}) can describe very
different physical scenarios. Here we wish to focus on the case of
a system composed by a pair of not directly interacting spins each
one coupled to every two-level component of a set of $N$ spins by
a physical mechanism representable by a XY Heisenberg exchange
like term. If we stipulate the absence of any internal coupling
within the set of $N$ spins, hereafter referred to as the spin
bath, and indicate by $A$ and $B$ the two spins of the preferred
couple, also called the central system, the hamiltonian model
(\ref{generalH}) assumes the following form
\begin{eqnarray}\label{HXY}
\nonumber
H_{XY}=&&\alpha_{A}(S_{x}^{A}\sum_{i=1}^NS_{x}^{i}+S_{y}^{A}\sum_{i=1}^NS_{y}^{i})
  +\\
  &&+\alpha_{B}(S_{x}^{B}\sum_{i=1}^NS_{x}^{i}+S_{y}^{B}\sum_{i=1}^NS_{y}^{i})
  \end{eqnarray}
It is worth noticing that specializing the hamiltonian model given
by eq. (\ref{generalH}) into the one expressed by eq. (\ref{HXY})
we are supposing that the spin $A$ and $B$ couples with any
element of the bath with a site-independent coupling constant
$\alpha_A$ or $\alpha_B$ respectively.

The interest toward this system, representing a generalization of
those systems reported in literature as spin star systems
\cite{XY1, spinstar}, arises from several considerations. First of
all, as we are going to show, it is exactly solvable so that in
this case we can explicitly know the functions $a(t)$, $b(t)$,...,
$e(t)$ appearing in the density matrix given by eq.
(\ref{rogenerica}). This circumstance allows us to explore
conditions for the emergence of and to investigate on the time
evolution of correlations get established within the pair $A$ and
$B$. A specific aspect of this study is in particular, the
possibility of controlling the rising of only classical or also
quantum correlations between the two not directly interacting
spins in the central subsystem. Another important point worth to
be emphasized is the fact that under appropriate initial
conditions, the concurrence appearing and evolving within the
$A-B$ subsystem may be traced back and interpreted in terms of
simple measurable quantities. As a consequence we are in condition
to suggest a way to measure the concurrence in laboratory without
being obliged to reconstruct, as usually required, the state of
the system.

Let's begin by observing that the possibility of exactly solving
the dynamics of the system described by (\ref{HXY}), is strictly
related to the existence of other constants of motion with respect
to the general model (\ref{generalH}).

Introducing the bath collective operators
  \begin{equation}\label{J}
  \overrightarrow{J}=\sum_{i=1}^N\overrightarrow{S}^{(i)}
  \end{equation}
   and
   \begin{equation}\label{Jpm}
   J_\pm=J_x\pm iJ_y
   \end{equation}
 and casting the interaction hamiltonian\ (\ref{HXY}) in the form
\begin{equation}\label{H_XYnuova}
H_{XY}=\alpha_{A}(S_{+}^{A}J_{-}+S_{-}^{A}J_{+})
+\alpha_{B}(S_{+}^{B}J_{-}+S_{-}^{B}J_{+})
\end{equation}
it is indeed easy to convince oneself that
\begin{equation}\label{CM J^2}
[H_{XY},J^2]=0
\end{equation}
\begin{equation}
[H_{XY},J_{int}^2]=0
\end{equation}
$\overrightarrow{J}_{int}$ being an intermediate angular momentum
resulting from the coupling of selected at will individual angular
momenta of the bath. For example defining
$\overrightarrow{J}_{i,j}=
\overrightarrow{S}^{(i)}+\overrightarrow{S}^{(j)}$, or
$\overrightarrow{J}_{i,j,k}=\overrightarrow{J}_{i,j}+\overrightarrow{S}^{(k)}$,
we deduce
\begin{equation}\label{CM Jij}
[H_{XY},J_{i,j}^2]=0\;\;\;\;\;[H_{XY},J_{i,j,k}^2]=0
\end{equation}

In order to make more evident how the existence of all these
constants of motion provides the possibility of exactly solving
the dynamics of the closed system constituted by the central one
$(A,B)$ and the spin bath, let's denote by $\{|J,M,\nu \rangle\}$
a bath coupled basis satisfying
\begin{eqnarray}
 &&J^2\vert J,M,\nu\rangle=J(J+1)\vert J,M,\nu\rangle \\
&&J_z\vert J,M,\nu\rangle=M\vert J,M,\nu\rangle
\end{eqnarray}
$\nu$ being  an integer index taking into account the degeneracy
with respect to the quantum numbers $J$ and $M$. As far as the
ordered basis of the $(A,B)$ subsystem given by eq. (\ref{standard
basis}), we introduce the new  more  compact notation
$\{|\sigma_z^A(n),\sigma_z^B(n)\rangle, \;\; n=1,..,4 \}$ putting
\begin{eqnarray} \label{standard basis (n)}
\nonumber && \vert \sigma_z^A(1),\sigma_z^B(1)\rangle\equiv \vert
1_A,1_B\rangle,\\ \nonumber && \vert
\sigma_z^A(2),\sigma_z^B(2)\rangle\equiv \vert 1_A,-1_B\rangle,\\
\nonumber && \vert \sigma_z^A(3),\sigma_z^B(3)\rangle\equiv \vert
-1_A,1_B\rangle, \\ && \vert
\sigma_z^A(4),\sigma_z^B(4)\rangle\equiv \vert -1_A,-1_B\rangle,
\end{eqnarray}
We can thus denote by $\vert J,M,\nu\rangle
\vert\sigma_z^A(n),\sigma_z^B(n)\rangle$$\equiv \vert
J,M,\nu,\sigma_z^A(n),\sigma_z^B(n)\rangle$ a generic state of the
basis of the Hilbert space of the total system obtained as
tensorial product of the two bases previously introduced.

It is immediate to convince oneself that the dynamical constraints
imposed by the constants of motion put into evidence at the
beginning of this section, allow us to claim that, starting from a
state $\vert \sigma_z^A(n),\sigma_z^B(n), J,M,\nu\rangle $, at a
generic time instant $t$ the state of the system will be at most a
linear superposition of four states as follows

\begin{eqnarray}\label{time evolution}
\nonumber &&\vert \psi(t)\rangle=\\
\nonumber &&A_{nJM}(t)\vert\sigma_z^A(n),\sigma_z^B(n),J,M,\nu\rangle+\\
 \nonumber
&&B_{nJM}(t)|-\sigma_z^A(n),-\sigma_z^B(n),J,M+\sigma_z^A(n)+\sigma_z^B(n),\nu\rangle+\\
\nonumber &&C_{nJM}(t)|\sigma_z^A(n),-\sigma_z^B(n),J,M+\sigma_z^B(n),\nu\rangle+\\
&&D_{nJM}(t)|-\sigma_z^A(n),\sigma_z^B(n),J,M+\sigma_z^A(n),\nu\rangle
\end{eqnarray}

We have demonstrated that it is possible to explicitly find the
exact form of the amplitudes $A_{nJM}(t)$, $B_{nJM}(t)$,
$C_{nJM}(t)$ and $D_{nJM}(t)$ appearing in eq. (\ref{time
evolution}) whatever the state of the coupled basis chosen as
initial state is. Here, for simplicity, we do not give their
analytical expressions also because, generally speaking, they are
highly involved. However it is important to stress that, knowing
the functions $A_{nJM}(t)$, ... $D_{nJM}(t)$ whatever $n,J$ and
$M$ are, we, at least in principle, can evaluate the temporal
evolution of the closed system from a generic initial state.

We now concentrate on a specific initial condition namely the one
wherein the two central spins are initially prepared one up and
the other one down whereas the bath is in its eigenstate of
minimum free energy:
\begin{equation}\label{initial state}
\vert \psi(0)\rangle=\vert
1^A,-1^B,\frac{N}{2},-\frac{N}{2},1\rangle
\end{equation}

We remark that the couple $J=\frac{N}{2}$, $M=-\frac{N}{2}$ is not
degenerate so that $\nu=1$ only. The time evolution of the system
starting from this initial condition is characterized by the fact
that the probability of finding the two spins of interest in the
state $\vert 1^A,1^B\rangle$ is zero at any time instant $t$. This
circumstance directly follows from the boundaries imposed to the
dynamics of the system by the conservation of the $z$ component of
the total angular momentum. Moreover it is possible to prove that
in the case under scrutiny the matrix element $c(t)$ is real. Thus
the reduced density matrix of the central system assumes the form
\begin{equation}\label{roAB}
\rho_{ij}(t) = \left( \begin{array}{cccc}
0 & 0 & 0 & 0 \\
0& b(t) & c(t) & 0 \\
0& c(t) & d(t)& 0\\
0& 0&  0& e(t)
\end{array} \right)
\end{equation}
where
\begin{equation} \label{b(t)AB}
b(t)=\frac{1}{(1+r^2)^2}[\cos(2\sqrt{N(1+r^2)}\alpha_At)+r^2]^2
\end{equation}
\begin{eqnarray} \label{c(t)AB}
\nonumber c(t)=&&
-\frac{r}{(1+r^2)^2}[\cos(2\sqrt{N(1+r^2)}\alpha_At)+r^2]\\&&[1-\cos(2\sqrt{N(1+r^2)}\alpha_At)]
\end{eqnarray}
\begin{equation} \label{d(t)AB}
d(t)=\frac{r^2}{(1+r^2)^2}[1-\cos(2\sqrt{N(1+r^2)}\alpha_At)]^2
\end{equation}
\begin{equation} \label{e(t)AB}
e(t)=\frac{1}{(1+r^2)}\sin(2\sqrt{N(1+r^2)}\alpha_At)^2
\end{equation}
In equations (\ref{b(t)AB})-(\ref{e(t)AB})
$r\equiv\frac{\alpha_B}{\alpha_A}$ measures the ratio of the two
coupling constants between each component of the central system
and the bath. Exploiting the results obtained in the previous
section expressed by eqs. (\ref{Cx}) and (\ref{conc Cx Cy}) we may
thus claim that in the case here analyzed, measuring the
covariance $C_{\sigma_x \sigma_x}$ corresponds to directly detect
the entanglement arisen between $A$ and $B$ being
\begin{equation}
C_{onc}= \vert C_{\sigma_x \sigma_x}\vert =2 \vert c(t)\vert
\end{equation}
as easily demonstrable looking at eqs.
(\ref{b(t)AB})-(\ref{e(t)AB}). In particular the circumstance that
we can explicitly solve the dynamics of the system provides the
possibility of knowing at a generic time instant $t$ the degree of
entanglement developed in the central system, starting from a
factorized condition, as measured by the concurrence function
$C_{onc}$.

Figure (1) displays the temporal behaviour of the concurrence
function $C_{onc}(t)$, obtained putting $N=100$, in correspondence
to three different values of $r$ namely $r=10^{-1}$, $r=1$ and
$r=10$ respectively.

\begin{figure}
\hspace{1.5cm}
\includegraphics[width=8 cm,height=4 cm]{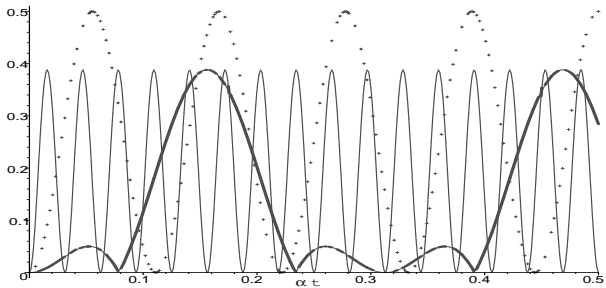}
\caption{ \footnotesize Concurrence function, $C_{onc}$, against
$\alpha_At$ in correspondence to $N=100$ and $r=1$ (point line),
$r=10$  and $r=10^{-1}$ (bold line) respectively.} \label{fig1}
\end{figure}

As clearly shown, varying the parameter $r$ the concurrence
function always manifests a periodic oscillatory behaviuor
immediately deducible taking into account the fact that, in the
case under scrutiny, $C_{onc}(t)=2\vert c(t)\vert$. In particular
we may state that, whatever $r$ is, there exist infinite values of
$t$ in correspondence of which $C_{onc}(t)=0$ meaning that the
central system is separable. It is worth emphasizing the
remarkable circumstance that when $r=1$ the matrix element $c(t)$
at any time instant $t$ assumes its maximum value compatible with
relation (\ref{Laundau condition}), that is $\vert
c(t)\vert=\sqrt{b(t)d(t)}$. In order to verify the occurrence of
such a saturation of the inequality (\ref{Laundau condition}) at
any $t$, it is enough to put $r=1$ in eqs.
(\ref{b(t)AB})-(\ref{d(t)AB}). On the other hand, it happens that
in correspondence to all the time instants such that
$C_{onc}(t)=0$ also the diagonal matrix element $e(t)$ vanishes.
This means that at these time instants all the conditions on the
density matrix of two spins defining a pure state  are verified
\cite{Landau}. Thus we may claim that when the two spins $A$ and
$B$ are equally coupled to the spin bath, namely $r=1$, the
condition $C_{onc}(t)=0$ not only reveals absence of entanglement
in the central system but also guarantees that such two spin
system is in a pure state. Looking indeed at eqs.
(\ref{b(t)AB})-(\ref{d(t)AB}) it is immediate to convince oneself
that the two spins are periodically found in the initial state
$\vert 1^A,-1^B\rangle$ or in the state $\vert -1^A,1^B\rangle$ in
which the role played by the two spins is exchanged. This
behaviour is clearly illustrated in figure (2) where we compare
the concurrence function with the temporal evolution of the
probability of finding the state $\vert 1^A,-1^B\rangle$  and that
of measuring the state obtained exchanging the two spins.

\begin{figure}
\hspace{1.5cm}
\includegraphics[width=8 cm,height=4 cm]{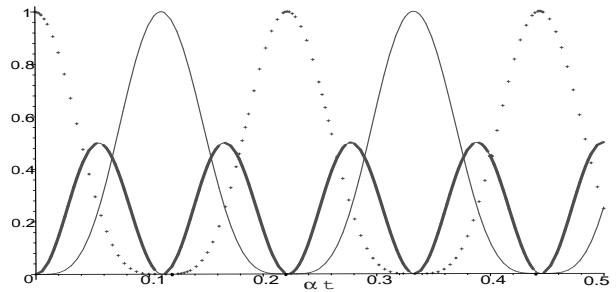}
\caption{ \footnotesize Concurrence function, $C_{onc}$,(bold
line), probability of finding the state $\vert -1^A,1^B\rangle$
(point line) and probability of finding the state $\vert
1^A,-1^B\rangle$ (continuous line), against $\alpha_At$ and in
correspondence to $N=100$ and $r=1$. } \label{fig2}
\end{figure}

Let's now examine how the system evolves starting from another
initial condition of experimental interest that is
\begin{equation}\label{initial condition 2}
\vert \psi(0)\rangle=\vert
1^A,1^B,\frac{N}{2},-\frac{N}{2},1\rangle
\end{equation}
obtained leaving once again the bath in its ground state and
preparing the preferred couple of spins in the state correspondent
to the maximum values of $S_z^{(AB)}$. It is possible to prove
that under this initial condition and assuming that the
Hamiltonian model (\ref{H_XYnuova}) is invariant  under the
permutation of $A$ with $B$ ($\alpha_A=\alpha_B\equiv \alpha$),
the central system develops correlations when the time goes on.
More in detail the temporal behaviour of $C_{\sigma_x\sigma_x}$ in
this case is given by
\begin{equation}\label{Csigmax2}
C_{\sigma_x\sigma_x}(t)=\frac{N}{3N-2}\sin^2{\omega t}
\end{equation}
where
\begin{equation}\label{omega}
\omega=2\sqrt{6N-4}\alpha
\end{equation}

The maximum amount of correlation between $\sigma_x^A$ and
$\sigma_x^B$ is obviously strictly related to the number of
external spins populating the bath but, however, is larger than
$\frac{1}{3}$ as immediately deducible by eq. (\ref{Csigmax2}). On
the other hand, evaluating the concurrence function exploiting the
results presented in section II yields
\begin{equation}\label{conc2}
C_{onc}(t)=\max{[0,f_N(x)]}
\end{equation}
where
\begin{eqnarray}\label{f_N}
 &&f_N(x)=\frac{2N\sqrt{2N(N-1)}-N(3N-2)}{(3N-2)^2}x^2+\\
\nonumber &&-\frac{2(N-2)\sqrt{2N(N-1)}}{(3N-2)^2}x+\\
\nonumber &&+ \frac{4(1-N)\sqrt{2N(N-1)}+N(3N-2)}{(3N-2)^2}
\end{eqnarray}
with $x\equiv\cos{\omega t}$ with $\omega$ given by eq.
(\ref{omega}). It is easy to verify that $f_N(x)$ is not positive
when $x\in [-1,1]$. Thus we may conclude that whatever $N$ is, the
concurrence function $C_{onc}(t)$ is equal to zero at any time
instant $t$. In other words, differently from the case before
analyzed, under the hypotheses before discussed the concurrence
function is equal to zero at any time instant $t$. Stated
differently the correlations occurring in the central system at a
generic time instant $t$, manifested throughout $C_{\sigma_x
\sigma_x}(t)\neq 0$, have only a classical origin and therefore we
say that such correlations are classical.

It is interesting to observe that this inability of the system to
generate entanglement between the spins $A$ and $B$ when the
system is prepared accordingly to eq. (\ref{initial condition 2}),
is overcome simply breaking the symmetry condition $r=1$. This
fact is clearly shown in figure (3) where we plot the two
functions $C_{onc}(t)$ and $C_{\sigma_x \sigma_x}(t)$ in
correspondence to $N=100$ and $r=10$.  Thus for $r\neq 1$ the
concurrence function remains zero for a given interval of time,
then abruptly increases and once again falls to zero. This
behaviour is then periodically recovered reaching maximum values
of $C_{onc}(t)$ of experimental interest. We wish to remark that
the evolution of the system in this case is characterized by
intervals of time during which the system develops only classical
correlations ($C_{onc}(t)=0$ and $C_{\sigma_x \sigma_x}(t)\neq 0$)
and intervals of time where quantum correlations occur
($C_{onc}(t)\neq 0$).

\begin{figure}
\hspace{1.5cm}
\includegraphics[width=8 cm,height=4 cm]{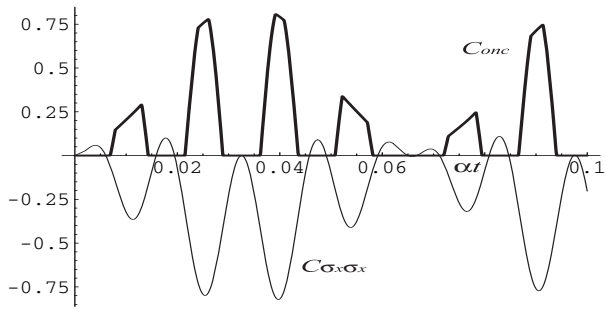}
\caption{ \footnotesize  Covariance function $C_{\sigma_x
\sigma_x}$ and concurrence $C_{onc}$ against $\alpha_At$, in
correspondence to $N=100$ and $r=10$} \label{fig3}
\end{figure}

\section{Conclusive remarks}

In this paper we have analyzed the quantum dynamics of a system of
$M$ Heisenberg coupled spins $\frac{1}{2}$ concentrating in
particular on a couple of them and looking for the time evolution
of entanglement developed between the two spins. We have
demonstrated that it is possible to establish a conceptual and
quantitative link between the concurrence function and easily
measurable quantities only requiring that at a generic time
instant $t$ the reduced density matrix describing the two spins of
interest can be put in the simple form given by eq.
(\ref{rogenerica}). This circumstance not only allows the
possibility of physically interpreting the concurrence function
introduced by Wootters as an useful and powerful mathematical tool
to estimate the degree of entanglement, but also suggests a direct
way to measure it. The quantitative relation given by eq.
(\ref{conc Cx Cy a e}) provides indeed the possibility of
measuring the entanglement without reconstructing the state of the
two spins. Examining in particular a specific physical system,
namely the generalized spin star system discussed in section III,
we have put into light that detecting the covariance function of
the two commuting observables $\sigma_x^A$ and $\sigma_x^B$
directly provides the entanglement evolution get established
between the two central spins. Exploiting the knowledge of the
exact expression of the density matrix of the system,
$\rho_{AB}(t)$, starting from an arbitrary initial condition, we
have disclosed a rich dynamics. In particular we have envisaged
physical conditions under which the two central spins are able to
develop classical correlations only being in this case
$C_{onc}(t)=0$ at any time instant $t$. At the same time we have
found that slightly changing some parameters characterizing the
system, quantum correlations appear.


\begin{thebibliography}{99}

\bibitem{Loss} D. Loss, D.P. Di Vincenzo, {\sl Phys. Rev. A}
{\bf 57}, 120 (1998).
\bibitem{Karbach} P. Karbach, J. Stolze, {\sl Phys. Rev. A}
{\bf 72}, 030301(R) (2005).
\bibitem{Dobrovitski} V.V. Dobrovitski, H. A. DeRaedt, M. I. Katsnelson, B. N. Harmon {\sl Phys. Rev. Lett.}
{\bf 90}, 210401 (2003).
\bibitem{Geza} Geza Toth, {\sl Phys. Rev. A}
{\bf 71}, 010301(R) (2005).
\bibitem{Cucchietti} F.M. Cucchietti, J. P. Paz, W. H. Zurek {\sl Phys. Rev. A}
{\bf 72}, 052113 (2005).
\bibitem{Burkard} G. Burkard, cond-mat/0409626.
\bibitem{Awschalom} D.D. Awschalom et al., { \sl Semiconductor
Spintronics and Quantum Computation}, Springer (Berlin 2002).
\bibitem{Imamoglu} A. Imamoglu et al, {\sl Phys. Rev. Lett.}
{\bf 83}, 4204 (1999).
\bibitem{XXX} M.C. Arnesen, S. Bose, V. Vedral {\sl Phys. Rev. Lett.}
{\bf 87}, 017901 (2001).
\bibitem{XXZ} Guo-Feng Zhang, Shu-Shen Li, {\sl Phys. Rev. A}
{\bf 72}, 034302 (2005).
\bibitem{XYZ1} L.-A. Wu, S. Bandyopadhyay, M. S. Sarandy, D. A. Lidar {\sl Phys. Rev. A}
{\bf 72}, 032309 (2005).
\bibitem{XYZ2} L. Zhou, H. S. Song, Y. Q. Guo, C. Li {\sl Phys. Rev. A}
{\bf 68}, 024301 (2003).
\bibitem{XY1}A. Hutton, S. Bose, {\sl Phys. Rev. A}
{\bf 69}, 042312 (2004).
\bibitem{XY2} S.D. Hamieh, M.I. Katsnelson, {\sl Phys. Rev. A}
{\bf 72}, 032316 (2005).
\bibitem{magnetism} Daniel C. Mattis, {\sl The theory of magnetism made
simple}, World Scientific Publishing Co. Pte. Ltd. (Singapore
2006)
\bibitem{Pratt}J.S. Pratt, {\sl Phys. Rev. Lett.}
{\bf 93}, 237205 (2004).
\bibitem{Landau} L.D. Landau, E.M. Lifsits, { \sl Quantum
Mechanics}
\bibitem{P-H1} A. Peres, {\sl Phys. Rev. Lett.}
{\bf 77}, 1413 (1996).
\bibitem{P-H2} M. Horodecki et al., {\sl Phys. Lett. A}
{\bf 223} (1996).
\bibitem{Wootters98} W. K. Wootters {\sl Phys. Rev. Lett.}
{\bf 80}, 2245-2248 (1998).
\bibitem{cavity1} Ting Yu, J.H. Eberly {\sl Phys. Rev. Lett.} {\bf 93},
140404 (2004).
\bibitem{cavity2} Ting Yu, J.H. Eberly, quant-ph/0503089 (2005)
\bibitem{cavity3}M. Yonac, T. Yu, J. H. Eberly, quant-ph/0602206
(2006).
\bibitem{spinstar} Heinz-Peter Breuer, D. Burgarth, F. Petruccione {\sl Phys. Rev. B.}
{\bf 70}, 045323 (2004).











\end{thebibliography}
\end{document}